\documentclass[12pt]{article}
\usepackage{amsmath}
\usepackage{blindtext} 

\usepackage[sc]{mathpazo} 
\usepackage[T1]{fontenc} 
\usepackage{microtype} 

\usepackage[english]{babel} 

\usepackage[hmarginratio=1:1,top=32mm,columnsep=20pt]{geometry} 
\usepackage[hang, small,labelfont=bf,up,textfont=it,up]{caption} 
\usepackage{booktabs} 

\usepackage{lettrine} 

\usepackage{enumitem} 
\setlist[itemize]{noitemsep} 

\usepackage{abstract} 
 
\usepackage{titlesec} 
\renewcommand\thesection{\Roman{section}} 
\renewcommand\thesubsection{\roman{subsection}} 
\titleformat{\section}[block]{\large\scshape\centering}{\thesection.}{1em}{} 
\titleformat{\subsection}[block]{\large}{\thesubsection.}{1em}{} 

\usepackage{fancyhdr} 
\usepackage{titling} 

\usepackage{hyperref} 

\pretitle{\begin{center}\Huge\bfseries} 
\posttitle{\end{center}} 
\title{Innermost stable circular orbits of a Kerr-like Metric with Quadrupole} 
\author{%
\textsc{Fabi\'an Chaverri Miranda} \\[1ex] 
\textsc{Francisco Frutos Alfaro} \\[1ex] 
\textsc{Pedro G\'omez Ovarez} \\[1ex] 
\textsc{Andree Oliva Mercado} \\[1ex]
\normalsize{School of Physics and 
Space Research Center of the University of Costa Rica}} 
\date{\today} 
\renewcommand{\maketitlehookd}{%
\begin{abstract}
\noindent 
The innermost stable circular orbit equation of a test particle is 
obtained for an approximate Kerr-like spacetime with quadrupole moment. 
We derived the effective potential for the radial coordinate by the 
Euler-Lagrange method. This equation can be employed to measure the mass 
quadrupole by observational means, because from this equation a quadratic 
polynomial for the quadrupole moment can be found. As expected, 
the limiting cases of this equation are found to be the known cases of 
Kerr and Schwarzschild.
\end{abstract}
}

\begin{document}

\maketitle


\section{Introduction}

In classical mechanics the orbit of a test particle around a massive object is 
arbitrary. This is because the effective potential is minimum, for any value of 
the angular momentum. Nevertheless, in general relativity this is not the case. 
The effective potential in the Schwarzschild metric has two extrema. 
When the angular momentum is minimum the two extrema become a single radius 
which describes the innermost stable circular orbit (ISCO) 
\cite{tsupko2016parameters}. 

Naturally, the rotation of the central body influences the motion around of a 
particle orbiting it, this is why the orbits around black holes differ between 
metrics \cite{jefremov2015innermost}. Another important feature of compact 
object is its quadrupole moment ($ q $) \cite{quevedo2016multipole}. 
It is expected that it would affect the orbits around compact objects. 
For many exact spacetimes, one cannot find analytical expresions for radii and 
frequencies of the ISCO. Moreover, Geodesics analysis is cumbersome for such 
metrics. It would be useful for studying wave emission and chaotic 
trajectories of particles around compact objects 
\cite{berti2004approximate,pachon2006realistic,sanabria2010innermost}.

Studying the ISCO of black holes is important, because it gives information 
about the spacetime near the black hole and its background geometry 
\cite{pradhan2012isco,quevedo2016multipole}. It is important to recall 
the no hair conjecture. It states that the geometry outside the black hole 
horizon (Kerr-Newman metric) is expressed only in terms of three parameters 
$ M $ (mass), $ a $ (rotation parameter) and $ e $ (charge). Nevertheless, 
for neutron stars this is not the case. Other parameters, such as deformation 
(mass quadrupole), and magnetic dipole should be taken into account, 
but they require the equation of state of the neutron star to completely 
describe the spacetime near them. However, we are concerned with compact 
objetcs with three parameters $ M $, $ a $ and $ q $
\cite{shibata1998innermost}.

As an example, in a Schwarzschild black hole the radius of the ISCO is 
$ r_{ISCO} = 6 M $. If the electric charge is zero, a Kerr type black hole is 
obtained from the no hair conjecture
\cite{pradhan2012isco,shibata1998innermost}. For the Kerr metric the radius of 
the ISCO depends on the direction of motion of the particle in comparison with 
the black hole. If the particle moves in the direction of the rotation of the 
black hole, then the radius becomes smaller $ r_{ISCO} = M $, if the particle 
moves counter rotation then the radius is bigger than Schwarzschild's and 
becomes $ r_{ISCO} = 9 M $ \cite{tsupko2016parameters}.
 
In this paper, we are interested in analysing the ISCO of a Kerr-like metric 
with mass quadrupole. The importance of this metric is that it reduces to 
the Kerr spacetime and to the Hartle-Thorne (HT) case for certain limits 
\cite{frutos2014approximate,frutos2013scirp}. In principle, it is possible 
to find an inner solution for the studied metric, by a match with a metric 
that already matches HT. Moreover, the ISCO equation was found for HT 
metric \cite{berti2004isco,kuantay2015geodesics}.

This Kerr-like metric was deduced from the Kerr metric, for this reason it is 
expected to obtain the known results for Kerr and Schwarzschild metrics for 
the variables energy, angular momentum and the ISCO radius 
\cite{frutos2016approximate,frutos2015approximate}.

We are also interested in the observational applications. Nowadays, there are 
no direct measurements of the quadrupole moment for compact objects, an 
analysis of the ISCO structure for this Kerr-like metric could give us a hint 
to derive $ q $ via observational methods.

This paper is organized as follows. The Kerr-like metric is introduced in 
section 2. A detailed calculation of the ISCO equation using the 
Euler-Lagrange method is in section 3. This method was developed by 
Chandrasekhar \cite{chandrasekhar1998mathematical}. In section 4, 
the ISCO equation is compared with the known solutions, Kerr and Schwarzchild 
black holes by means of a REDUCE program. The summary and discussion of the 
results are presented in section 5.


\section{The Kerr-like Metric}

This metric describes the spacetime of a massive, rotating, deformed object. 
It has three parameters, the mass of the object, $ M $ , the rotation 
parameter, $ a $ and the quadrupole parameter, $ q $. 
It is an approximate solution of the Einstein field equations and is given by

\begin{align}
d s^{2} & = g_{tt} d t^{2} + 2 g_{t\phi} d t d \phi + g_{rr} d r^{2} 
+ g_{\theta\theta} d \theta^{2} + g_{\phi\phi} d \phi^{2} ,
\end{align}

\noindent
where the components of the metric are

\begin{align}
\label{met1} 
g_{tt} & = - \dfrac{{\rm e}^{-2\psi}}{\rho^{2}}\left[\Delta 
- a^{2} \sin^{2}{\theta} \right] \nonumber \\
g_{t\phi} & = -\dfrac{2Jr}{\rho^{2}} \sin^{2}{\theta} \nonumber \\
g_{rr} & = \rho^{2} \dfrac{{\rm e}^{2\chi}}{\Delta} \\
g_{\theta\theta} & = \rho^{2} {\rm e}^{2\chi} \nonumber \\
g_{\phi\phi} & = \dfrac{{\rm e}^{2\psi}}{\rho^{2}}
\left[\left(r^{2} + a^{2}\right)^{2} - a^{2} \Delta \sin^{2}{\theta} \right] 
\sin^{2}{\theta} , \nonumber \\
\end{align}

\noindent
with $ J = M a $,  $ \rho^{2} = r^{2} - a^{2}\cos^{2}{\theta} $ and 
$ \Delta = r^{2} - 2 M r + a^{2} $. The exponents $ \psi $ and $ \chi $ are 
given by

\begin{align}
\psi & = \dfrac{q}{r^{3}}P_{2} + 3 \dfrac{M q}{r^{4}}P_{2} \\
\chi & = \dfrac{q}{r^{3}}P_{2} + \dfrac{1}{3} \dfrac{M q}{r^{4}} 
\left(- 1 + 5 P_{2} + 5 P_{2}^{2}\right) \nonumber 
+ \dfrac{1}{9} \dfrac{q^{2}}{r^{6}}\left(2 - 6 P_{2} - 21 P_{2}^{2} 
+ 25 P_{2}^{3}\right) \nonumber
\end{align}

\noindent
where $ P_{2} = (3 \cos^{2}{\theta} - 1)/2 $. As limiting cases this spacetime 
contains the Kerr ($ q = 0 $) and the Schwarzschild metrics ($ q = a = 0 $).


\newpage

\section{Deriving the ISCO}

The method, we are using, was devised by Chandrasekhar 
\cite{chandrasekhar1998mathematical}. The Lagrangian is given by

\begin{align}
L & = \dfrac{\mu}{2}\dfrac{ds^{2}}{d\lambda^{2}} \nonumber \\
  & = \dfrac{\mu}{2}(g_{tt}\dot{t}^{2} + 2 g_{t\phi} \dot{t} \dot{\phi} 
+ g_{rr}\dot{r}^{2} + g_{\theta\theta}\dot{\theta}^{2} + g_{\phi\phi}\dot{\phi}^{2}) .
\end{align}

The dot over the variables $ t $, $ r $, $ \theta $ and $ \phi $ means 
derivative with respect to $ \lambda $. To determine the ISCO of a test 
particle in the plane, one sets $ \dot{\theta} = 0 $ and 
$ \theta = {\pi}/{2} $. This leaves the Lagrangian as follows

\begin{align}
L = \dfrac{\mu}{2}\left(g_{tt} \dot{t}^{2} + 2 g_{t\phi} \dot{t} \dot{\phi} 
+ g_{rr} \dot{r}^{2} + g_{\phi\phi} \dot{\phi}^{2} \right) ,  
\end{align}

\noindent
where the components of the metric becomes

\begin{align}
\label{met2} 
g_{tt} & = \dfrac{e^{- 2 \psi'}}{r^2} \left[2Mr-r^{2} \right] 
\nonumber \\
g_{t\phi} & = - \dfrac{2J}{r} \nonumber \\
g_{rr} & = \dfrac{r^{2} {\rm e}^{2\chi'}}{\Delta}  \\
g_{\theta\theta} & = r^{2} {\rm e}^{2\chi'} \nonumber \\
g_{\phi\phi} &=\dfrac{{\rm e}^{2\psi'}}{r^2}\left[r^{4} + 2M r a^{2} 
+ r^{2} a^{2}\right] \nonumber 
\end{align}

\noindent
with the exponents $ \psi $ and $ \chi$ are reduced to

\begin{align}
\psi' & = - \frac{1}{2} \dfrac{q}{r^{3}} - \frac{3}{2} \dfrac{M q}{r^{4}} 
\label{exp1} \\
\chi' & = - \frac{1}{2} \dfrac{q}{r^{3}} - \frac{3}{4} \dfrac{M q}{r^{4}} 
- \frac{3}{8} \dfrac{q^{2}}{r^{6}} \label{exp2} \nonumber
\end{align}

The momenta of the three remaining variables are

\begin{align}
p_{t} &=\mu\left(g_{tt}\dot{t} + g_{t\phi}\dot{\phi}\right) = - E \nonumber \\
p_{r} &=\mu g_{rr}\dot{r} \\
p_{\phi} &=\mu\left(g_{t\phi}\dot{t}+g_{\phi\phi}\dot{\phi}\right) = L_{z} \nonumber
\end{align}

\noindent
where $ \mu = 1 $, $ b^{2} = - g_{tt} g_{\phi\phi} + g_{t\phi}^{2} $, 
$ E $ and $ L_{z} $ are constants that represent the energy and the angular 
momentum. $ \dot{t} $ and $ \dot{\phi} $ are solved

\begin{align}
\dot{t} & = \dfrac{1}{b^{2}} \left(E g_{\phi\phi} + L_{z} g_{t\phi} \right) \\
\dot{\phi} & = - \dfrac{1}{b^{2}} \left(L_{z} g_{tt} + E g_{t\phi} \right) . 
\nonumber
\end{align}

The Hamiltonian is

\begin{align}
H & = \dfrac{1}{2}\left(- E \dot{t} + g_{rr} \dot{r}^{2} 
+ L_{z}\dot{\phi}\right) = \varepsilon \\
& = \dfrac{1}{2} \bigg[ - \dfrac{1}{b^{2}}
\left(E^{2} g_{\phi\phi} + 2 E L_{z} g_{t\phi} + L_{z}^{2} g_{tt}\right) 
+ g_{rr} \dot{r}^{2} \bigg] \nonumber 
\end{align}

\noindent
The effective potential $ V_{ef} $ is then

\begin{align}
V_{ef} & = - \dfrac{2\varepsilon}{g_{rr}} 
- \dfrac{1}{g_{rr} b^{2}}(E^{2} g_{\phi\phi} 
+ 2 E L_{z} g_{t\phi} + L_{z}^{2} g_{tt}) \label{Vef}
\end{align}

Using $ u = 1/r $ and the equations \ref{exp1} to \ref{met2}, then 
the effective potential in \ref{Vef} can be reduced to

\begin{align}
V_{ef} & = - 2\varepsilon \bigg(1 - 2 M u + a^{2} u^{2} + q u^{3} 
- \frac{1}{2}{M q u^{4}} + \frac{5}{4} {q^{2} u^{6}} \bigg) \nonumber \\
& + L_{z}^{2} \left(u^{2} + 2 q u^{5} + \frac{1}{2} {M q u^{6}} 
+\frac{11}{4} {q^{2} u^{8}} \right) - 2 M u^{3}(L_{z} - E a)^{2} \nonumber \\
& - E^{2} \left(1 + a^{2} u^{2} - \frac{3}{4} {M q u^{4}} 
+ \frac{3}{4} {q^{2} u^{6}} \right) \label{peff} 
\end{align}

The latter expression has to be differentiated twice to find the values of 
$ r $ where the orbit is stable.

\begin{align}
\dfrac{dV_{ef}}{du} & = - 2 \varepsilon \bigg(- 2 M + 2 a^{2} u 
+ 3 q u^{2} - 2 M q u^{3} + \frac{15}{2} {q^{2} u^{5}} \bigg) \nonumber \\
& + L_{z}^{2} \left(2 u + 10 q u^{4} + 3 M q u^{5} + 22 q^{2} u^{7} \right) 
- 6 M u^{2} \left(L_{z} - E a \right)^{2} \nonumber \\
& - E^{2} \left(2 a^{2} u - 6 M q u^{3} + \frac{9}{2} {q^{2} u^{5}} 
\right) \label{dpeff}
\end{align}

\begin{align} 
\dfrac{d^{2}V_{ef}}{du^{2}} 
& = - 2 \varepsilon \left(2 a^{2} + 6 q u - 6 M q u^{2} 
+ \frac{75}{2} {q^{2} u^{4}} \right) \nonumber \\
& + L_{z}^{2} \left(2 + 40 q u^{3} + 15 M q u^{4} + 154 q^{2} u^{6} \right) 
- 12 M u \left(L_{z} - E a \right)^{2} \nonumber \\
& - E^{2} \left(2 a^{2} - 18 M q u^{2} + \frac{45}{2} {q^{2} u^{4}} \right) 
\label{d2peff}
\end{align}

Now, we rewrite these equations using $ x = a E - L_{z} $ as follows

\begin{align}
{\tilde V}_{ef} & = V_{ef} [1 - 2 q u^3 + (2 q u^3)^2  \nonumber \\
& = \varepsilon \bigg(2 - 4 M u + 2 a^2 u^2 - 2 q u^3 
+ 7 M q u^4 + \dfrac{13}{2} q^2 u^6 \bigg) \nonumber \\
& + E^2 \bigg(- 1 + 2 q u^3 + \dfrac{3}{2} M q u^4 
- \dfrac{19}{4} q^2 u^6 \bigg) - 2 E x a u^2 \nonumber \\
& + x^2 u^2 \bigg(1 - 2 M u + \dfrac{9}{2} M q u^4 + \dfrac{11}{4} q^2 u^6 
\bigg) = 0 , \label{pefftecho} \\
\dfrac{{d} {\tilde V}_{ef}}{{d} {u}} & = 
\dfrac{{d} {V}_{ef}}{{d} {u}} [1 - 5 q u^3 + (5 q u^3)^2]  \nonumber \\
& = \varepsilon\big(- 4 M + 4 a^2 u + 6 q u^2 + 16 M q u^3 
- 15 q^2 u^5\big) \nonumber \\
& + E^2 q u^3 \left(6 M - \dfrac{9}{2} q u^2 \right) 
- 4 E x a u \nonumber \\
& + x^2 u \big(2 - 6 M u + 33 M q u^4 + 22 q^2 u^6 \big) = 0 , 
\label{dpefftecho} \\
\dfrac{{d}^2 {\tilde V}_{ef}}{{d} {u}^2} & = 
\dfrac{{d}^2 V_{ef}}{{d} {u}^2} [1 - 20 q u^3 + (20 q u^3)^2] \nonumber \\
& =\varepsilon (4 a^2 + 12 q u - 12 M q u^2 - 165 q^2 u^4) \nonumber \\
& + E^2 q u^2 \left(18 M - \dfrac{45}{2} q u^2 \right) - 4 E x a \nonumber \\
& + x^2 \big(2 - 12 M u + 255 M q u^4 + 154 q^2 u^6\big) = 0 , 
\label{d2pefftecho}
\end{align}

\noindent
where the expressions are set to zero, because we are interested in 
determining the ISCO equation.

From \ref{pefftecho} and \ref{dpefftecho}, $ E^2 $ is found

\begin{align}
E^2 & = \varepsilon(2 - 2 M u - q u^3 - 8 M q u^4 + 7 q^2 u^6) \nonumber \\
& + x^2 u^3 \left(M - 10 M q u^3 - \frac{33}{4} q^2 u^5 \right) . \label{E2}
\end{align}

A fourth order polynomial for $ x $ is obtained from \ref{dpefftecho} and 
\ref{E2}

\begin{align} 
{\cal A} x^4 + 2 {\cal B} x^2 + {\cal C} = 0 , 
\end{align} 

\noindent
where

\begin{align} 
{\cal A} & = u^2 (1 - 6 M u + 9 M^2 u^2 - 4 M a^2 u^3 
+ 33 M q u^4 + 22 q^2 u^6) \nonumber \\
{\cal B} & = - 2\varepsilon u \bigg(M + (a^2 - 6 M^2) u 
\nonumber \\
&+ \left(M a^2 u^2 - \frac{3}{2} q \right) u^2 - \frac{5}{2} M q u^3 
+ 6 q^2 u^5\bigg) \nonumber \\
{\cal C} & = \varepsilon^{2} \big(4 M^2 - 8 M a^2 u + 4 a^4 u^2 
- 12 M q u^2 + 9 q^2 u^4 \big).
\end{align}

The solution for $ x^2 $ is given by

\begin{align}
x^2 & =\frac{2\varepsilon}{u Z_{\mp}} 
\bigg[\left(a \sqrt{u} \pm \sqrt{M} \right)^2 
- \frac{3}{2} q u^2 - 7 M q u^3 + 6 q^2 u^5 \bigg] \label{x2} ,
\end{align}

\noindent
with

\begin{align}
Z_{\pm} & = \left(1 - 3 M u + \frac{33}{2} M q u^4 + 11 q^2 u^6 \right) 
\pm 2 a u \sqrt{M u} .
\end{align}

Inserting \ref{x2} in \ref{E2} one finds $ E^2 $  

\begin{align}
E^2 & = \frac{2\varepsilon}{Z_{\mp}}
\bigg[(1 - 2 M u) \left(1 - 2 M u \pm 2 a u \sqrt{M u} \right) \label{E2R} \\
& + \left(a^2 M - \frac{1}{2} q \right) u^3 + \frac{25}{2} M q u^4 
+ \frac{29}{2} q^2 u^6 \bigg] \nonumber
\end{align}

Substituting \ref{E2R} and \ref{x2} in \ref{dpeff}, $ L_{z}^{2} $ is determined

\begin{align}
L_z^2 & =\frac{2\varepsilon u}{{\cal A}} \bigg[M - 3 M^2 u 
+ \left(2 M a^2 - \frac{3}{2} q \right) u^2 
+ \left(6 M^2 a^2 - \frac{5}{2} M q \right) u^3 \nonumber \\
& + M \left(a^4 - 12 M^2 a^2\right) u^4 + \left(5 M^2 a^4 + 6 q^2\right) u^5 
\nonumber \\
& \pm 2 M a u \sqrt{M u} \big(a^4 u^4 - 2 M a^2 u^3 
+ 4 a^2 u^2 - 6 M u + 3\big) \bigg], \label{Lz2}
\end{align}

\noindent
where $ {\cal A} = u^2 Z_{+} Z_{-} $.

Finally, substituting \ref{E2R}, \ref{Lz2} and \ref{x2} in \ref{d2peff} 
and changing $ u = 1/r $, the ISCO equation is found

\begin{align}
{\cal P} & = M r^5 - 9 M^2 r^4 
+ 3 \left(6 M^3 - M a^2 + \frac{1}{2} q \right) r^3 \nonumber \\
& - \left(7 M^2 a^2 - \frac{29}{2} M q \right) r^2 
- \frac{33}{2} q^2 \pm 6 M a r \sqrt{M r} \Delta = 0 . \label{pisco}
\end{align}


\section{Comparing with known metrics}

Now, we proceeded to analyze the limiting cases for the possible $ r $ values. 
For this analysis we also used a REDUCE program which finds the solutions for 
equations such as the one obtained in \ref{pisco}.

The first limiting case is when $ q = a = 0 $ which reduces to the known 
Schwarzschild metric case ($ r = 6 M $). For Schwarzschild, the relation found 
is the following

\begin{align}
{\cal{P}}_{Sch} & = M r^{5}- 9 M^{2} r^{4} + 18 M^{3} r^{3} 
= M r^{3} \left(r - 6 M \right) \left(r - 3 M \right) = 0 \label{piscosch}
\end{align}

Clearly, the Schwarzschild case is contained in \ref{piscosch}. It is also 
found the values of the energy and angular momentum, reducing \ref{x2} and 
\ref{E2} and using $ r = 6 M $ and $ \varepsilon = 1/2 $.

\begin{align}
E & = \dfrac{1 - 2 M u}{\sqrt{1 - 3 M u}} \nonumber \\
& = \sqrt{\dfrac{8}{9}} \label{VE} \\
L_{z} & = \sqrt{\dfrac{M}{u\left(1 - 3 M u \right)}} \nonumber \\
& = 2 \sqrt{3} M \label{VLz}
\end{align}

As seen in \ref{VE} and \ref{VLz} both values are the ones found by 
Chandrasekhar \cite{chandrasekhar1998mathematical}. 

The other important case is Kerr, reducing \ref{pisco} to Kerr means that 
$ q = 0 $. The ISCO equation for the Kerr metric found by Chandrasekhar and 
Pradhan \cite{chandrasekhar1998mathematical,pradhan2012isco} is

\begin{align}
r^{2} - 6 M r \mp 8 a \sqrt{M r} - 3 a^{2} & = 0
\label{kerrchand} ,
\end{align}

\noindent
Squaring the last expression, one gets

\begin{align}
r^{4} - 12 M r^{3} + 6 (6 M^{2} - a^{2}) r^{2} - 28 M a^{2} r + 9 a^{4} & = 0 
\label{kerrlisco} .
\end{align}

\noindent
The simplification of \ref{pisco} is

\begin{align}
{\cal{P}}_{Kerr} & = M r^{5} - 9 M^{2} r^{4} 
+ 3 \left(6 M^{3} -M a^{2} \right) r^{3} 
- 7 M^{2} a^{2} r^{2} \pm 6 M a r \sqrt{M r}\Delta = 0 . \label{piscokerr}
\end{align}

\noindent
From last expression, after squaring, we get

\begin{align}
(r^{4} - 12 M r^{3} + 6 (6 M^{2} - a^{2}) r^{2} - 28 M a^{2} r + 9 a^{4}) 
& \times & \nonumber \\
(r^{3} - 6 M r^{2} + 9 M^{2} r - 4 M a^{2}) & = 0 \label{piscokerr2}
\end{align}

Obviously equation \ref{kerrchand} is contained in \ref{piscokerr}, 
giving us the solutions for the Kerr case. The energy and angular momentum are

\begin{align}
E & = \sqrt{\frac{1}{Z_{\mp}}} \bigg(1 - 2 M u \mp a u\sqrt{M u} \bigg), \\
x & = - \dfrac{a \sqrt{u} \pm \sqrt{M}}{\sqrt{u Z_{\mp}}}, \nonumber \\
L_{z} &= \mp \sqrt{\dfrac{M}{u Z_{\mp}}}\left(a^{2} u^{2} + 1 
\pm 2 a u \sqrt{M u}\right).
\end{align}

These values for $ E $ and $ L_z $ are exactly the same as the ones 
determined by Chandrasekhar \cite{chandrasekhar1998mathematical} which 
validates the original equations \ref{x2}, \ref{E2} and \ref{Lz2}.


\section{Discussion}

We have successfully found the ISCO equation for a Kerr-like metric with 
quadru\-pole, that reduces to the known equations for Kerr ($ q = 0 $) and 
Schwarzchild ($ q = a = 0 $) metrics. 

Graphical analysis concerning the dependence of the ISCO radius, energy and 
angular momentum with the mass, rotation parameter, and quadrupole parameter 
will be discussed in a future article.

An important feature of the ISCO equation, is that it is quadratic in $ q $, 
therefore it is easily solvable given the values of $ a $, $ M $ and 
$ r_{ISCO} $. This means that, given this model, indirect measurement of the 
quadrupole parameter of a compact object can be completely done from 
observational data regarding the mass, rotational parameter and ISCO radius of 
the object.

Further applications, including analysis of observational data, as well as 
extending the ISCO equation for the inclusion of charged compact objects, 
will be done in future papers.




\end{document}